\documentclass[fleqn,usenatbib]{mnras}

\usepackage{newtxtext,newtxmath}
\usepackage[T1]{fontenc}

\DeclareRobustCommand{\VAN}[3]{#2}
\let\VANthebibliography\thebibliography
\def\thebibliography{\DeclareRobustCommand{\VAN}[3]{##3}\VANthebibliography}

\DeclareRobustCommand{\DE}[3]{#2}
\let\DEthebibliography\thebibliography
\def\thebibliography{\DeclareRobustCommand{\DE}[3]{##3}\DEthebibliography}

\DeclareRobustCommand{\DEN}[3]{#2}
\let\DENthebibliography\thebibliography
\def\thebibliography{\DeclareRobustCommand{\DEN}[3]{##3}\DENthebibliography}

\usepackage{graphicx}	% Including figure files
\usepackage{amsmath}	% Advanced maths commands
\usepackage{booktabs}
\usepackage{float}
\usepackage{threeparttable}
\usepackage{longtable}
\usepackage{pdflscape}
\usepackage{breqn}
\usepackage{tikz,xcolor,hyperref}

\definecolor{lime}{HTML}{A6CE39}
\DeclareRobustCommand{\orcidicon}{%
    \begin{tikzpicture}
    \draw[lime, fill=lime] (0,0) 
    circle [radius=0.16] 
    node[white] {{\fontfamily{qag}\selectfont \tiny ID}};
    \draw[white, fill=white] (-0.0625,0.095) 
    circle [radius=0.007];
    \end{tikzpicture}
    \hspace{-2mm}
}

\newcommand{\orcidmpr}{\href{https://orcid.org/0000-0001-9164-2882}{\orcidicon}}
\newcommand{\orcidnad}{\href{https://orcid.org/0000-0003-4842-8834}{\orcidicon}}

\title[Tungsten in barium stars]{Tungsten in barium stars}

\author[M. Roriz et al.]{
M.~P.~Roriz\orcidmpr,$^{1}$\thanks{E-mail: michelle@on.br}
M.~Lugaro,$^{2,3,4}$
S.~Junqueira,$^{1}$
C.~Sneden,$^{5}$
N.~A.~Drake\orcidnad,$^{1,6}$
and C.~B.~Pereira$^{1}$\\
\\
$^{1}$Observat\'orio Nacional/MCTI, Rua General Jos\'e Cristino, 77, 20921-400, Rio de Janeiro, Brazil\\
$^{2}$ Konkoly Observatory, Research Centre for Astronomy and Earth Sciences, Eötvös Loránd Research Network (ELKH), H-1121 Budapest, \\Konkoly Thege M. \'ut 15-17, Hungary\\
$^{3}$ ELTE E\"{o}tv\"{o}s Lor\'and University, Institute of Physics, Budapest 1117, P\'azm\'any P\'eter s\'et\'any 1/A, Hungary\\
$^{4}$ School of Physics and Astronomy, Monash University, VIC 3800, Australia\\
$^{5}$ Department of Astronomy and McDonald Observatory, The University of Texas, Austin, TX 78712, USA\\
$^{6}$ Laboratory of Observational Astrophysics, Saint Petersburg State University, Universitetski pr. 28, 198504, Saint Petersburg, Russia
}

\date{Accepted XXX. Received YYY; in original form ZZZ}

\pubyear{2015}

\begin{document}
\label{firstpage}
\pagerange{\pageref{firstpage}--\pageref{lastpage}}
\maketitle

\begin{abstract}
Classical barium stars are red giants that received from their evolved binary companions material exposed to the \textit{slow} neutron-capture nucleosynthesis, i.e., the $s$-process. Such a mechanism is expected to have taken place in the interiors of Thermally-Pulsing Asymptotic Giant Branch (TP-AGB) stars. As post-interacting binaries, barium stars figure as powerful tracers of the $s$-process nucleosynthesis, evolution of binary systems, and mechanisms of mass transfer. The present study is the fourth in a series of high-resolution spectroscopic analyses on a sample of 180 barium stars, for which we report tungsten (W, $Z=74$) abundances. The abundances were derived from synthetic spectrum computations of the W\,{\sc i} absorption features at 4\,843.8~\AA\ and 5\,224.7~\AA. We were able to extract abundances for 94 stars; the measured [W/Fe] ratios range from $\sim0.0$ to $2.0$ dex, increasing with decreasing of metallicity. We noticed that in the plane [W/Fe] versus [$s$/Fe] barium stars follow the same trend observed in post-AGB stars. The observational data were also compared with predictions of the {\sc fruity} and Monash AGB nucleosynthesis models. These expect values between $-0.20$ and $+0.10$ dex for the [W/hs] ratios, whereas a larger spread is observed in the program stars, with [W/hs] ranging from $-0.40$ to $+0.60$ dex. The stars with high [W/hs] ratios may represent evidence for the operation of the intermediate neuron-capture process at metallicities close to solar.
\end{abstract}


%%%%%%%%%%%%%%%%%%%%%%%%%%%%%%%%%%%%%%%%%%%%%%%%%%
%%%%%%%%%%%%%%%%% BODY OF PAPER %%%%%%%%%%%%%%%%%%

\section{Introduction}

A large fraction of elements beyond the iron-peak ($Z>30$) is produced in the interiors of Thermally-Pulsing Asymptotic Giant Branch stars \citep[TP-AGB;][]{galino1998, straniero2006, karakas2014} through the \textit{slow} neutron-capture mechanism \citep[$s$-process;][]{kappeler2011, lugaro2023}. From convective mixing episodes, the newly synthesized nuclei are brought to the stellar surface, changing the chemical composition of the star. Subsequently, the enriched material is ejected to the interstellar medium via stellar mass-loss events. In addition to $s$-elements, AGB stars can also play an important role in the Galactic chemical evolution of C, N, and F \citep[e.g.,][]{kobayashi2020}. Generally, these objects are observed as S-type stars. Indeed, the detection of the radioactive $^{99}$Tc on their atmospheres \citep[][]{merrill1952}{}{} provided an unambiguous observational evidence that the $s$-process nucleosynthesis is on-going within these stars.

On the other hand, observations also show that about 50\% of S-type stars do not show Tc absorption features in their spectra \citep[see][]{vaneck2022}{}{}. These outlier stars are probably first ascent giants, therefore unable to internally produce the heavy elements observed in their envelopes. In fact, they are thought to be members of post-interacting binary systems, where mass transfer took place and originated their chemical peculiarities \citep[e.g.,][]{jorissen1998, jorissen2019}{}{}. In that framework, the observed stars received from their evolved companion the $s$-processed material. Thus, they are referred to as \textit{extrinsic}, instead of being \textit{intrinsic} S stars. The $s$-rich sub-class of the Carbon-Enhanced Metal-Poor stars (i.e., CEMP-$s$; \citealt[][]{beers2005}), giant CH stars \citep[][]{keenan1942}, and classical barium stars \citep[][]{bidelman1951} are representative of such extrinsic stars.

The classical barium stars are the warmest extrinsic $s$-rich stars; they are G/K spectral-type giants with effective temperatures ranging from $4\,000$ to $6\,000$~K. The abnormal strengthening of the Ba\,{\sc ii} and Sr\,{\sc ii} absorption lines along with the relatively intense CH, CN, and C$_{2}$ molecular bands observed in their spectra posed a challenge to early stellar evolution models. \citet[][]{mcclure1980} first reported radial velocity variations in these systems, and shed light on the binary nature of barium stars. Over the years, such a scenario has been confirmed and extensively explored from data of radial velocity monitoring programs \citep[][]{jorissen1988, jorissen1998, jorissen2019, escorza2019, escorza2023}. As post-interacting binary systems, the orbital elements of barium stars have provided valuable observational constraints to binary evolution models, as well as clues on mass transfer mechanisms.

Beginning with the first quantitative abundance analysis of the barium star HD~46407 \citep{burbidge1957}, many studies have found moderated carbon enhancements ($\textrm{[C/Fe]}\approx0.30$)\footnote{Throughout this study, the standard spectroscopy notation is used, namely ${\rm [A/B]}=\log(N_{\rm A}/N_{\rm B})_{\star} -\log(N_{\rm A}/N_{\rm B})_{\odot}$, where the $\odot$ symbol refers to solar values, together with $\log\epsilon({\rm A})=\log(N_{\rm A}/N_{\rm H})+12$.} and $s$-process average abundances that can reach values $\textrm{[\textit{s}/Fe]}>1.0$ in barium stars. Although a lower limit in the [$s$/Fe] index to categorize an object as a barium star is not well established, \citet[][]{decastro2016} suggested $\textrm{[\textit{s}/Fe]}>0.25$. Irrespective of the degree of C and $s$-enhancement observed in various barium stars, they figure as powerful $s$-process tracers of their cooler counterparts (i.e., intrinsic S-stars), providing strong constraints to the nucleosynthesis models \citep[e.g.,][]{allen2006a, allen2006b, smiljanic2007, pereira2011, yang2016, decastro2016, karinkuzhi2018, cseh2018, shejeelammal2020, roriz2021a, roriz2021b, cseh2022}. To trace back the envelope abundances of a barium star to the former TP-AGB companion -- now quietly orbiting it as a dim white dwarf -- it is important to explore all available elemental abundances on its atmosphere.

The present paper is the fourth in a series of studies started by \citet[][hereafter Paper~I]{decastro2016}, who conducted a chemical and kinematic analysis for a large sample of barium giant stars, based on high-resolution spectroscopic data. In Paper~I, $\sim180$ targets were subjected to statistical analysis. Later, Rb abundances were extracted for these stars in \citet[][hereafter Paper~II]{roriz2021a}, since Rb is a key element to constraint the neutron source of the $s$-process \citep[see][]{vanraai2012}. In paper~II, we found $\textrm{[Rb/Zr]}<0$ for these systems. In the light of theoretical $s$-process nucleosynthesis models, this is expected when $^{13}$C acts as the main neutron source of the $s$-process \citep[][]{galino1998}. In other words, Paper~II provided an additional evidence for the low-mass nature (i.e., $\lesssim3~\textrm{M}_{\odot}$) of the former AGB stars that contaminated the envelopes of the observed barium stars. In \citet[][hereafter Paper~III]{roriz2021b}, we re-derived the La abundances previously reported in Paper~I, and extracted new abundances for other neutron-capture ($n$-) elements, namely Sr, Nb, Mo, Ru, Sm, and Eu. Most importantly, for the elements Nb, Mo, and Ru, which are nuclei between the first (Sr, Y, Zr) and second (Ba, La, Ce) $s$-process peaks, the observations were at odds with nucleosynthesis models. We found abundances systematically higher than theoretical predictions from the $s$-process, which led us to suggest alternative nucleosynthesis paths to interpret the observations \citep[see also][]{cseh2022, denhartogh2023}{}{}. 

In the present study, we aim to extend the abundance pattern of barium stars by evaluating tungsten (W, $Z=74$) abundances in our sample of 180 barium giants. As far as we are aware, W stellar abundances (or upper limits) have been reported in the literature just for a few sources: the barium star HD~46407 \citep[][]{burbidge1957}, 8 post-AGB stars \citep[][]{reyniers2003, reyniers2004, desmedt2012, vanaarle2013, desmedt2016}, and 5 metal-poor ($\textrm{[Fe/H]}\lesssim-1.50$) stars \citep[][]{siqueira2013, roederer2014, roederer2016, roederer2022}. More recently, \citet[][]{roriz2023}{}{} derived W abundances in 1 new barium giant (HE~0457-1805), 1 new CH star (HE~1255-2324), and 1 probable CH star (HE~2207-1746). W is an element located between the second and third (Pb) $s$-process peaks, and its solar abundance is predicted to have a contribution of $62\%$ from the $s$-process \citep[][]{bisterzo2014}. The abundance of W produced by the $s$-process mostly follows that of the second peak at Ba and La, and can be used as a diagnostic of the neutron flux \citep[see, e.g.,][]{reyniers2003}{}{}. For example, \citet[][]{lugaro2015}{}{} demonstrated that the relatively high W and low Pb abundances in post-AGB stars cannot be explained by the $s$-process, which calls for the intermediate ($i$-) neutron-capture process \citep[][]{cowan1977}{}{} to be responsible for the observed pattern.

The paper is organized as follows. In Section~\ref{sec:targets}, we provide a summary of the target stars. In Section~\ref{sec:analysis}, we describe the approach in deriving W abundances and estimating their uncertainties. We discuss the novel results in Section~\ref{sec:discussion}, and analyse the data set in the light of the $s$-process nucleosynthesis models in Section~\ref{sec:comparison}. Finally, in Section~\ref{sec:conclusions}, we draw our conclusions.

\section{Program stars}\label{sec:targets}

The programme stars comprise 166 sources of Paper~I, 11 metal-rich ($\textrm{[Fe/H]}\geq0.10$) barium stars previously analyzed by \citet[][]{pereira2011}, 2 barium stars identified by \citet[][]{katime2013} in the open cluster NGC~5822 (\#2 and \#201), and the barium star HD~10613 from \citet[][]{pereira2009}. Therefore, a total of 180 targets. The atmospheric parameters adopted here are the same previously adopted in Papers~II and III. For completeness, they are shown in Table~\ref{app:tab:abd}. In summary, these stars were observed between the years 1999 and 2010, and their high-resolution spectra were acquired with the Fiber-fed Extended Range Optical Spectrograph \citep[FEROS;][]{kaufer1999} attached to the 1.52 and 2.2 m ESO telescopes at La Silla (Chile). The instrument has a spectral coverage within the range from $\sim3\,900$ to $9\,200$~\AA\ with a resolving power $R=\lambda/\Delta\lambda\sim48\,000$.

\section{Abundance analysis}\label{sec:analysis}

\subsection{Abundance derivation}

\begin{figure}
    \centering
    \includegraphics{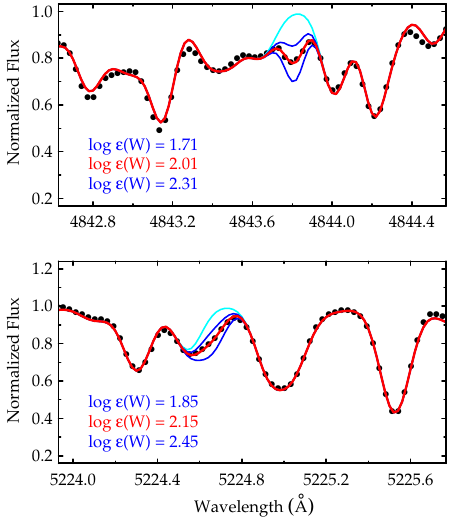}
    \caption{A portion of the spectral region around the W\,{\sc i} lines at 4\,843.8~\AA\ (top panel) and 5\,224.6~\AA\ (bottom panel). The black dots represent the observed spectrum of the star HD~107541, and the curves are synthetic spectra computed for different W abundances, as indicated in each panel. The best fit syntheses are shown in red, whereas the blue curves are spectra computed for $\Delta\log\epsilon(\textrm{W})=\pm0.3$ around the adopted solution. In cyan, we show theoretical spectra computed without W contribution.}
    \label{fig:hd107541_syn}
\end{figure}

As previously mentioned, W abundances were reported so far only for 17 stars. In addition to the limited number of useful diagnostic lines, spectral blends make difficult the task of deriving the chemical abundance of W. Previous works, based on near-ultraviolet spectra of metal-poor stars, reported W abundances from eight W\,{\sc\,ii} lines, mainly the transitions at 2\,088.204~\AA\ and 2\,118.875~\AA\ \citep[see][]{siqueira2013, roederer2014, roederer2016, roederer2022}. In the optical window, the W\,{\sc\,i} line at 5\,053.280~\AA\ \citep[][]{vanaarle2013} and the W\,{\sc\,ii} line at 5\,104.432~\AA\ \citep[][]{reyniers2003, reyniers2004, desmedt2012, desmedt2016} were detected in the spectra of post-AGB stars. Recently, also in the optical domain, \citet[][]{roriz2023}{}{} used two absorption features of W\,{\sc\,i} centered at 4\,843.810~\AA\ and 5\,224.661~\AA\ to derive W abundances for the stars HE~0457-1805, HE~1255-2324, and HE~2207-1746 already mentioned in the Introduction.

To extract W abundances in the program stars, we have followed the same approach carried out by \citet[][]{roriz2023}. In other words, we fitted synthetic spectra to the W\,{\sc\,i} features at 4\,843.810~\AA\ and 5\,224.661~\AA. To model the theoretical spectra, we run the \textit{synth} drive of the radiative-transfer code {\sc moog} \citep[][]{sneden1973, sneden2012}, which assumes a plane-parallel stellar atmosphere and the Local Thermodynamic Equilibrium (LTE) conditions. To compute the synthetic spectra, {\sc moog} needs as inputs an atmospheric model and a linelist containing the lab data of relevant transitions in that spectral region. We have also adopted here the 1D plane-parallel model atmospheres of \citet[][]{kurucz1993}, as done in previous papers. For the W lines, we adopted the line parameters provided by the Vienna Atomic Line Database ({\sc vald}; \citealt[][]{ryabchikova2015}). In Appendix~\ref{app:data}, we provide the abundances derived from each line, as well as our adopted values. The transition at 4\,843.810~\AA\ has an excitation potential $\chi=0.412$~eV and $\log$~\textit{gf}~$=-1.54$, while the other at 5\,224.661~\AA\ has $\chi=0.599$~eV and $\log$~\textit{gf}~$=-1.70$. Tungsten has five stable isotopes, among which only the $^{183}$W presents a non-zero nuclear spin ($I=1/2$); however, hyperfine structure data are not available for that nuclide. 

In Figure~\ref{fig:hd107541_syn}, we show as an example the spectral regions close to the two W\,{\sc i} lines observed for the star HD~107541, as well as synthetic spectra computed for different W abundances. Note that the W\,{\sc i} line at $4\,843.8$~\AA\ is clearly present in the spectrum of this star, as shown in the upper panel of the figure. On the other hand, as seen in the lower panel of Figure~\ref{fig:hd107541_syn}, the absorption at $5\,224.6$~\AA\ is a blended feature; however, to fit the observation, the contribution of W\,{\sc i} at 5\,224.661~\AA\ in the spectral synthesis computations is necessary. In Table~\ref{app:tab:abd}, we provide the abundances derived from each line for the target stars, as well as the adopted values in this work. Furthermore, an inspection of Table~\ref{app:tab:abd} shows that abundances derived from the two W\,{\sc i} lines agree reasonably well, which confirms the reliability of the tungsten determination in our sample of barium stars. Tungsten abundances were derived for a total of 94 out the 180 stars analyzed here. All of these, except one (HD~88562), exhibited in their spectra the line at 4\,843.810~\AA, whereas the line at 5\,224.661~\AA\ was detected in 73 of them. For the others 86 program stars, the two W\,{\sc i} lines were absent or weakly detectable, so that abundances were not evaluated. Temperature and the degree of $s$-enrichment are two relevant factors for the appearance of the W\,{\sc i} lines in the spectra of the target stars. This follows directly from the Saha-Boltzmann formulas, as well as the fact that the line intensity grows with the increase of atmospheric W abundance.

For the chemical species considered in Papers~I, II, and III, recommendations of \citet[][]{grevesse1998} for the solar photosphere abundances were systematically used. Such assumption implies in a mean difference of $-0.04\pm0.04$ with respect to the updated values of \citet[][]{asplund2009}. On the other hand, as far as W is concerned, its solar photosphere abundance reported by \citeauthor[][]{grevesse1998}{}{}, $\log\epsilon(\textrm{W})_{\odot}=1.11$, is flagged as less accurate, and a much lower meteorite value is recommended by these authors, namely $\log\epsilon(\textrm{W})_{\odot}=0.65$. The recommendations of \citeauthor[][]{asplund2009}{}{} for the solar photosphere and meteorite abundances are $\log\epsilon(\textrm{W})_{\odot}=0.85$ and $0.65$, respectively, and also present an offset, but in a lesser extent than in \citeauthor[][]{grevesse1998}{}{}. Thus, in view of that issue, and to perform a fair comparison with theoretical predictions (Section~\ref{sec:comparison}), which are computed from meteorite abundances, we have adopted in this work the meteorite value of 0.65, the same value adopted in the nucleosynthesis models. For consistency, we have also normalized the data compiled from the literature to that value.

\subsection{Uncertainty estimates}\label{sec:uncertainty}

\begin{table*}
\centering
\caption[]{Abundance uncertainty estimates for the template stars HD~114678, HD~29370, and HD~211954, performed for the elements iron and tungsten. Columns from 2 to 6 give the variations introduced in abundances owing by changes in $T_{\rm eff}$, $\log g$, $\xi$, [Fe/H], and equivalent width measurements ($W_\lambda$), respectively. Column 7 gives the random errors. Finally, the composed uncertainties are shown in Column 8.}
\label{tab:err}
   \begin{threeparttable}
	\begin{tabular}{lccccccc}
		\toprule
  	Species & $\Delta T_{\textrm{eff}}$ & $\Delta \log{g}$ & $\Delta \xi$ & $\Delta$[Fe/H] & $\Delta W_{\lambda}^{\textrm{(a)}}$ & $\sigma_{\textrm{ran}}^{\textrm{(b)}}$ & $\sqrt{\Sigma \sigma^{2}}$ \\
        \midrule
        \multicolumn{8}{c}{Group 1 -- HD~114678}\\ \midrule  
  	             & ($+100$ K) & ($+0.2$ dex) & ($+0.3$ km/s) & ($+0.1$ dex) & ($+3$ m\AA) &         &      \\\midrule
        Fe\,{\sc i}   & $+0.10$    & $0.00$       &  $-0.11$      &  $-0.01$     & $+0.06$     & 0.02    & 0.16 \\
        W\,{\sc i}    & $+0.23$    & $0.00$       &  $0.00$       &  $0.00$      & -           & $0.20$ & 0.30\\
        \midrule[1.0pt]
        \multicolumn{8}{c}{Group 2 -- HD~29370}\\
		\midrule
  	             & ($+100$ K) & ($+0.2$ dex) & ($+0.3$ km/s) & ($+0.1$ dex) & ($+3$ m\AA) &     &      \\\midrule
        Fe\,{\sc i}   & $+0.08$    & $+0.01$      &  $-0.13$      &  $0.00$      & $+0.05$ & $0.02$  & 0.16 \\
        W\,{\sc i}    & $+0.20$    & $+0.03$      &  $0.00$       &  $+0.03$     & -       & $0.15$ & 0.25 \\
        \midrule[1.0pt]
        \multicolumn{8}{c}{Group 3 -- HD~211954} \\
		\midrule
  	             & ($+90$ K) & ($+0.2$ dex) & ($+0.3$ km/s) & ($+0.1$ dex) & ($+3$ m\AA) &         &      \\\midrule
        Fe\,{\sc i}   & $+0.03$   & $+0.02$      &  $-0.16$      &  $+0.02$     & $+0.06$     & $0.03$  & 0.18 \\
        W\,{\sc i}    & $+0.18$   & $+0.03$      &  $-0.20$      &  $+0.03$     & -           & $0.18$ & 0.33\\
		\bottomrule
	\end{tabular}
 \textbf{Notes:} (a) Evaluated only for the iron lines, from data of Paper~I. (b) For the iron lines, $\sigma_{\textrm{ran}}$ is evaluated as $\sigma_{\textrm{obs}}/\sqrt{N}$, where $\sigma_{\textrm{obs}}$ is the standard deviation and $N$ is the number of iron lines used, from data of Paper~I. 
     \end{threeparttable}
\end{table*}

To evaluate the uncertainties associated to the W abundances, we have grouped the stars into three ranges of effective temperatures. In Group 1, we selected stars with temperatures between 5\,000 and 5\,400~K; in Group 2, stars with 4\,700 and 4\,950~K; in Group 3, stars with 4\,100 and 4\,600~K. In our previous studies, the stars BD~$-14^{\circ}$2678, HD~119185, and HD~130255 were taken as templates of Groups 1, 2, and 3, respectively. However, as the W lines were weakly detectable in these targets, they were replaced by the stars HD~114678, HD~29370, and HD~211954, respectively.

By varying the atmospheric parameters, i.e., effective temperature ($T_{\textrm{eff}}$), surface gravity ($\log~g$), microturbulent velocity ($\xi$), and metallicity ([Fe/H]), we computed the corresponding changes introduced in $\log\epsilon(\textrm{Fe})$ and $\log\epsilon(\textrm{W})$. Additionally, to take into account the random error ($\sigma_{\textrm{ran}}$) due to continuum uncertainties, we computed the minimal abundance variations for which a clear visual difference is seen between the synthetic and observed spectra. The total uncertainties in $\log\epsilon(\textrm{Fe})$ and $\log\epsilon(\textrm{W})$ for the template star were estimated by adding quadratically the changes introduced in abundance and extracting the root square. In this approach, we assume that the errors are independent. The total abundance uncertainties are shown in the last column of Table~\ref{tab:err}. These values were applied to the objects of the respective sub-sample, as shown in Table~\ref{app:tab:abd}. Finally, for the [W/Fe] ratios, the uncertainties were estimated according to the relationship $\sigma_{\textrm{[W/Fe]}}^{2}=\sigma_{\textrm{W}}^{2}+\sigma_{\textrm{Fe}}^{2}$.

\section{Discussion}\label{sec:discussion}

\begin{figure}
    \centering
    \includegraphics{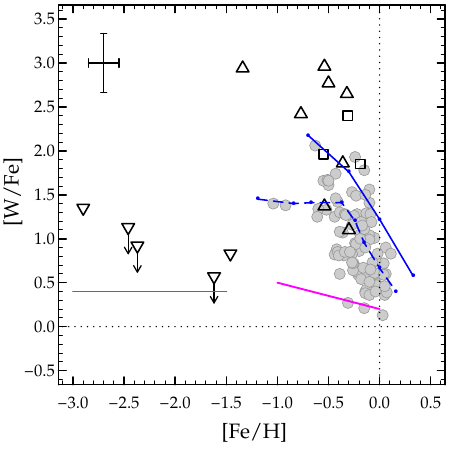}
    \caption{Tungsten-to-iron abundance ratios observed in barium giant stars (gray dots) as a function of metallicity. For the sake of clarity, only a typical error bar of data set is shown. Black symbols are data reported in the literature for metal-poor stars (inverted triangles; from \citealt[][]{siqueira2013, roederer2014, roederer2016, roederer2022}{}{}), post-AGB stars (up triangles; from \citealt[][]{reyniers2003, reyniers2004, desmedt2012, vanaarle2013, desmedt2016}{}{}), and other chemically peculiar stars (squares; from \citealt[][]{roriz2023}{}{}). As far as we are aware, no other stellar W abundances have been reported in the literature in addition to these data. The magenta lines mimic the GCE models computed by \citet[][]{kobayashi2020}, as seen in their Figure 32. Blue curves are some examples of theoretical expectations of the $s$-process for TP-AGB stars of 3.0 M$_{\odot}$ computed by the Monash group (solid line) and {\sc fruity} database (dashed line); these models will be discussed in Section~\ref{sec:comparison}.}
    \label{fig:feh_wfe}
\end{figure}

In Figure~\ref{fig:feh_wfe}, we plot the [W/Fe] ratios observed in our barium giants (gray dots) as a function of metallicity. We found W abundances spanning within the interval $0.0\lesssim\textrm{[W/Fe]}\lesssim+2.0$. Data for [W/Fe] available in the literature for post-AGB stars, metal-poor stars, and other chemically peculiar systems are also added in the same plot as black symbols; these are properly identified in the figure caption. As Figure~\ref{fig:feh_wfe} shows, five W abundances are available for metal-poor stars, among which 3 are upper limits for the [W/Fe] ratios. The data do not evidence any trend for $\textrm{[Fe/H]}\lesssim-1.50$. Additionally, in this metallicity range, $s$-process nucleosynthesis is not expected to play a significant role. Indeed, the Galactic chemical evolution (GCE) models of \citet[][]{kobayashi2020}, performed for elements from C to U, predict $\textrm{[W/Fe]}\sim0.40$, if contributions from magneto-rotational supernovae are taken into account (see Figure~\ref{fig:feh_wfe}); otherwise, values close to $\textrm{[W/Fe]}\sim-0.70$ are predicted.

For $\textrm{[Fe/H]}>-1.0$, the data shown in Figure~\ref{fig:feh_wfe} reflect directly the $s$-process nucleosynthesis. This is noticeable by the [W/Fe] ratios decreasing with increasing [Fe/H], a typical feature of the $s$-process \citep[][]{busso2001, cseh2018}{}{}. Post-AGB stars typically show the highest [W/Fe] values. For normal field stars in this metallicity range, W abundances are not available. Examples of theoretical [W/Fe] ratios, expected by AGB nucleosynthesis models of 3~M$_{\odot}$ and able to reach $\textrm{[\textit{s}/Fe]}>0.25$, are also shown in Figure~\ref{fig:feh_wfe} (blue curves). They are plotted in this figure only to illustrate how much W these models are able to produce. They should not be compared directly with our observational data set, since the material received by the barium stars is further diluted on their convective envelopes. These models will be discussed in detail in Section~\ref{sec:comparison}. 

\begin{figure}
    \centering
    \includegraphics{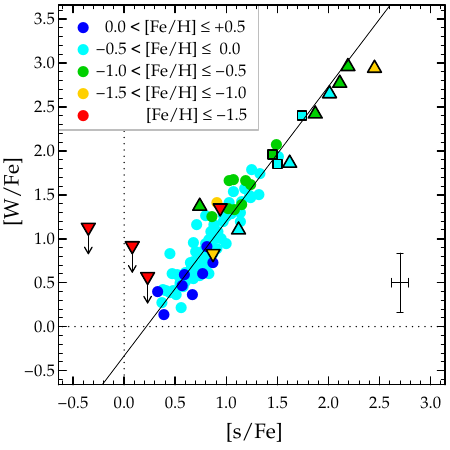}
    \caption{[W/Fe] ratios against [$s$/Fe] for the programme stars (color dots), along with data from the literature. The symbols have the same meaning as in Figure~\ref{fig:feh_wfe}, however, as shown in the top of panel, we adopted here a code color in order to identify the different metallicity regimes of the stars. For the sake of clarity, only a typical error bar is shown. The black straight line is a linear fit of data observed in barium giants.}
    \label{fig:sfe_wfe}
\end{figure}

In Figure~\ref{fig:sfe_wfe}, the [W/Fe] ratios are plotted as a function of the $s$-process mean abundance, denoted by the [$s$/Fe] index. This index is evaluated as the average of [X/Fe] ratios derived for the Sr, Y, Zr, La, Ce, and Nd -- elements of the first and second $s$-process peaks. The [$s$/Fe] values are given in the last column of Table~\ref{app:tab:abd} and lie in the range of $+0.23\leq$~[$s$/Fe]~$\leq+1.59$\footnote{In Paper~III, we included in the computations of [$s$/Fe] ratios abundances of Rb, as well as abundances of the elements Nb, Mo, and Ru, which are located between the first and second $s$-process peaks. Here we have adopted only elements of the first and second $s$-process peaks.}. As previously mentioned, there is no consensus concerning the lower limit for the [$s$/Fe] index to consider an object as a barium star. We have adopted here the condition [$s$/Fe]~$\geq0.25$ dex assumed in Papers~I, II, and III. We have also observed that for barium stars with [$s$/Fe]~$<0.40$ dex, the two W\,{\sc i} lines are not detectable in their spectra (see also Table~\ref{app:tab:abd}). Figure~\ref{fig:sfe_wfe} clearly demonstrates the strong correlation between [W/Fe] and [$s$/Fe] (as well as [Fe/H]) observed in our barium stars. Such a correlation confirms that W is produced within the former polluting TP-AGB stars, since the $s$-process enrichment levels are accompanied by higher W abundances. A least square fitting of the data set provides a relationship given by $\textrm{[W/Fe]}=(1.49\pm0.06)\times\textrm{[\textit{s}/Fe]}-(0.41\pm0.06)$. Interestingly, we note that the sources previously investigated in the literature fall very closely the linear fit, except for the metal-poor stars HD~108317, HD~128279, and HD~94028, for which only upper limits could be determined \citep[][]{roederer2014, roederer2016}.

\section{Comparison to nucleosynthesis models}\label{sec:comparison}

\begin{figure}
    \centering
    \includegraphics{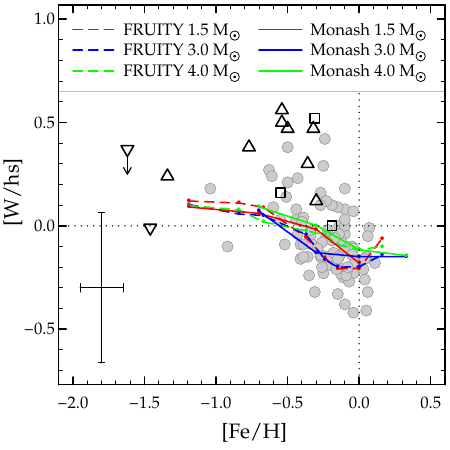}
    \caption{[W/hs] ratios observed in barium giant stars as a function of metallicity. The observations are compared to predictions from the Monash (continuous line) and {\sc fruity} (dashed lines) nucleosynthesis models for AGB of initial masses of 1.5, 3.0, and 4.0~M$_{\odot}$. Data for metal-poor stars and post-AGB stars are also shown in the plot. Symbols have the same meaning as in Figure~\ref{fig:feh_wfe}.}
    \label{fig:feh_whs}
\end{figure}

AGB stars experience thermal instabilities and dredge-up episodes which, in turn, favour nucleosynthesis via $s$-process \citep[][]{galino1998, busso1999, busso2001, straniero2006, karakas2014, lugaro2023}. Between the He- and H-burning shells, which are alternately activated in the TP-AGB phase, there is a small intershell region ($\sim10^{-2}$~M$_{\odot}$) rich in He and C where the $s$-process nucleosynthesis takes place. For low-mass stars, the $^{13}\textrm{C}(\alpha,n)^{16}\textrm{O}$ reaction provides the major neutron reservoir of the $s$-process. This reaction is activated at $T\sim10^{8}$~K in radiative conditions, during the H-burning shell (interpulse period) \citep[][]{straniero1995, galino1998}. The other neutron source, the $^{22}\textrm{Ne}(\alpha,n)^{25}\textrm{Mg}$ reaction, does not play a relevant role as neutron source for this mass range. It is marginally activated at $T\sim3\times10^{8}$~K, during the brief and recurrent He-burning shell episodes (thermal-pulse; TP), which temporarily extinguish the H fusion. Additionally, the CNO cycle in the H-burning shell does not leave in the intershell the required $^{13}$C amounts to drive the $s$-process \citep[][]{galino1998}. Therefore, to model the nucleosynthesis of heavy elements, a partial mixing zone is assumed in the top layers of intershell, allowing the penetration of protons from the convective envelope during the third dredge-up (TDU) episodes. These protons are combined with the $^{12}$C nuclei present there to form a $^{13}$C pocket, where $^{13}$C burns in the next interpulse. However, the exact mechanism that leads to the formation of the $^{13}$C pocket in the intershell is not well understood, which introduce the major uncertainty source in predictions (see, e.g., \citealt[][]{karakas2014}{}{}; \citealt[][]{lugaro2023}{}{}).

Theoretical predictions for the $s$-process yields produced in TP-AGB stars for a wide range of masses ($1.0<M/\textrm{M}_{\odot}<8.0$) and metallicities ($-1.2\lesssim\textrm{[Fe/H]}\lesssim+0.3$) are tabulated and made publicly available by the INAF group \citep[][]{cristallo2009, cristallo2011, cristallo2015}, through the {\sc fruity}\footnote{FUll-Network Repository of Updated Isotopic Tables \& Yields, online at: \url{http://fruity.oa-teramo.inaf.it/}} database, and the Monash group \citep[][]{fishlock2014, karakas2016, karakas2018}. Since that metallicity range covers the interval observed in barium stars, we examine our observations in light of these two sets of $s$-process nucleosynthesis models. Because the abundance profiles observed in barium stars have evidenced the most likely low-mass ($M\lesssim3.0~\textrm{M}_{\odot}$) nature of their former polluting TP-AGB stars (e.g., \citealt[][]{karinkuzhi2018, cseh2018, cseh2022, shejeelammal2020}{}{}; Paper~II; Paper~III), we concentrate our attention in (non-rotating) models of masses equal to 1.5, 3.0, and 4.0~$\textrm{M}_{\odot}$. 

A careful comparison between {\sc fruity} and Monash models, highlighting their similarities and differences, is provided by \citet[][]{karakas2016}{}{}, to which we refer the reader. Generally speaking, in the {\sc fruity} models, the $^{13}$C pocket is generated via a time-dependent convective overshoot implementation \citep[][]{cristallo2009}. In the Monash models, instead, a post-processing is performed to compute the detailed nucleosynthesis. In this step, the $^{13}$C pocket is formed from a parametric-approach, in which a mass of protons ($M_{\textrm{mix}}$) is artificially inserted in the top layers of the intershell during the TDU \citep[][]{karakas2016}. Additionally, different physical and nuclear inputs are used in the {\sc fruity} and Monash models. We have selected here only the models with the standard choice of the mass of $^{13}$C pocket, able to reach $\textrm{[\textit{s}/Fe]}>0.25$. We then compare our observations with predictions computed at the stellar surface at the end of the AGB evolution.

In Figure~\ref{fig:feh_wfe}, models of $3.0~\textrm{M}_{\odot}$ are shown as example. Note that they are able to produce [W/Fe] ratios increasing from $\sim0.5$ to $\sim2.2$ with decreasing metallicities. As we previously mentioned, the predicted [W/Fe] ratios cannot be directly compared to observations, since the models do not account for the mass transfer and the further dilution of the $s$-processed material on the atmospheres of barium stars. In any case, the predictions behave similarly to the observations, since dilution has the effect of lowering the predicted W abundances but without changing the shape of the distribution \citep[see, e.g.,][]{cseh2022}{}{}. In order to eliminate dilution effects and compare theoretical predictions with observations, we evaluated ratios between elements from different $s$-process peaks. In Figure~\ref{fig:feh_whs}, we plot the ratios $\textrm{[W/hs]}=\textrm{[W/Fe]}-\textrm{[hs/Fe]}$ as a function of metallicity, where [hs/Fe] denotes the average abundance of the elements La, Ce, and Nd, belonging to the second $s$-process peak. As shown in this figure, the [W/hs] ratios observed in barium stars span between $-0.40$ and $+0.60$, and most of the stars ($\sim90\%$ of the targets for which W abundances were evaluated) exhibit $\textrm{[W/hs]}<0$. The $s$-process model results are mostly controlled by the relatively well-known neutron-capture cross sections of the isotopes involved. Predicted trends show mild variations, of up to +0.3 dex when decreasing [Fe/H], and are not able to cover the full observational spread. Some such spread may be attributed to the observational error bars, however, there are a few data points that show significant excesses of W. These may be interpreted as a signature of the $i$-process \citep[][]{lugaro2015}{}{}.

\section{Conclusions}\label{sec:conclusions}
We have reported new observational data of tungsten abundances for a sample of 180 barium giant stars. So far, abundances (or upper limits) of this element are found in the literature only for 17 targets. Based on the LTE assumptions, W abundances were derived for 94 barium stars via spectral synthesis of two W\,{\sc i} absorption features centered at 4\,843.8~\AA\ and 5\,224.7~\AA. For the remaining 86 stars of the sample, these two W\,{\sc i} lines were either absent or weakly detectable in their spectra. We found [W/Fe] ratios spanning from $\sim0.0$ to $2.0$ dex, increasing for lower metallicities and with typical uncertainties of the order of $\pm0.30$ dex. As discussed, such abundances cannot be explained as a consequence of the GCE. In fact, GCE models predict no more than $+0.50$ dex \citep[][]{kobayashi2020}{}{}. The strong correlation between the [W/Fe] and [$s$/Fe] ratios observed in the atmospheres of program barium stars evidences that the W was produced by the $s$-process nucleosynthesis. We also noticed that in the plane [W/Fe] versus [$s$/Fe] barium stars follow the same trend observed in post-AGB stars, as shown in Figure~\ref{fig:sfe_wfe}. Finally, the [W/hs] ratios were compared with predictions from the {\sc fruity} and Monash nucleosynthesis models, which expect $-0.20<\textrm{[W/hs]}<+0.10$. We found that the $s$-process models are able to reproduce the bulk of observations, although the spread of the data set is greater than predictions, as seen in Figure~\ref{fig:feh_whs}. Higher [W/hs] than predicted may be the signature of the $i$-process, especially in connection to lower than expected [Pb/hs] abundances. Further data on the other elements located just before or at the third $s$-process peak will shed more light on this possibility and may provide evidence for the existence of the $i$-process at metallicities close to solar.

\section*{Acknowledgements}

This work has been developed under a fellowship of the PCI Program of the Ministry of Science, Technology and Innovation - MCTI, financed by the Brazilian National Council of Research - CNPq, through the grant 300438/2024-9. M.L. acknowledges the support of the Hungarian Academy of Sciences via the Lend\"ulet grant LP2023-10. C. S. thanks the U.S. National Science Foundation for support under grant AST 1616040. N.A.D. acknowledges Funda\c{c}\~ao de Amparo \`a Pesquisa do Estado do Rio de Janeiro - FAPERJ, Rio de Janeiro, Brazil, for grant E-26/203.847/2022. The authors would like to thank the referee for providing comments that improved the readability of the manuscript. This research has made use of NASA’s Astrophysics Data System Bibliographic Services and the . This work has made use of the {\sc vald} database, operated at Uppsala University, the Institute of Astronomy RAS in Moscow, and the University of Vienna.

%%%%%%%%%%%%%%%%%%%%%%%%%%%%%%%%%%%%%%%%%%%%%%%%%%
\section*{Data Availability}
The data underlying this article are available in the article.

\bibliographystyle{mnras}
\bibliography{references} % if your bibtex file is called example.bib

%%%%%%%%%%%%%%%%%%%%%%%%%%%%%%%%%%%%%%%%%%%%%%%%%%
%%%%%%%%%%%%%%%%% APPENDICES %%%%%%%%%%%%%%%%%%%%%
\appendix

\section{Abundance data}\label{app:data}

\clearpage

\onecolumn

\begin{longtable}{lcccccccc}
\caption{Tungsten elemental abundances derived for the barium giant stars of the program. The targets are listed in the first column. Effective temperature, surface gravity, and metallicity are shown in Columns 2, 3, and 4, respectively. The logarithmic W abundances derived from the W\,{\sc i} lines at 4\,843~\AA\ and 5\,224~\AA\ are shown in Columns 5 and 6. The adopted logarithmic abundances and the [W/Fe] ratios, as well as their uncertainties, computed according to Section~\ref{sec:uncertainty}, are listed in Columns 7 and 8, respectively. Finally, we show in Column 9 the averaged $s$-process abundances, given by the mean of the [X/Fe] ratios for the elements Sr, Y, Zr, La, Ce, and Nd, reported in Papers~I and III.}\label{app:tab:abd}\\
    \toprule
        Star  & $T_{\textrm{eff}}$ & $\log g$      & [Fe/H] & \multicolumn{3}{c}{$\log\epsilon(\textrm{W})$}          & [W/Fe] & [$s$/Fe] \\[0.1 cm]
                                                                         \cline{5-7}
              &        (K)         & (cm.s$^{-2}$) &        & W\,{\sc i} 4\,843~\AA & W\,{\sc i} 5\,224~\AA & Adopted &        &          \\
        
    \endfirsthead
    
    \multicolumn{9}{c}%
    {{\tablename\ \thetable{} -- Continued}} \\
    
    \toprule
        Star  & $T_{\textrm{eff}}$ & $\log g$      & [Fe/H] & \multicolumn{3}{c}{$\log\epsilon(\textrm{W})$}          & [W/Fe] & [$s$/Fe] \\[0.1 cm]
                                                                         \cline{5-7}
              &        (K)         & (cm.s$^{-2}$) &        & W\,{\sc i} 4\,843~\AA & W\,{\sc i} 5\,224~\AA & Adopted &        &          \\
        
    \midrule
    
    \endhead
    
    \midrule
    
    \multicolumn{9}{r}{{\textit{Continued on next page}}} \\
    
    \endfoot
    
    \bottomrule

    \endlastfoot
    
    \midrule
BD$-08^{\circ}$3194   &  4900 & 3.0 & $-0.10\pm 0.16$ & 2.01 & 1.91 & $1.96\pm 0.25$ & $+1.41\pm 0.30$ & $+1.09\pm 0.11$ \\
BD$-09^{\circ}$4337   &  4800 & 2.6 & $-0.24\pm 0.21$ & 2.41 & 2.26 & $2.34\pm 0.25$ & $+1.93\pm 0.30$ & $+1.42\pm 0.11$ \\
BD$-14^{\circ}$2678   &  5200 & 3.1 & $+0.01\pm 0.12$ &  ... &  ... &            ... &             ... & $+0.87\pm 0.10$ \\
CD$-27^{\circ}$2233   &  4700 & 2.4 & $-0.25\pm 0.18$ & 1.11 & 1.11 & $1.11\pm 0.25$ & $+0.71\pm 0.30$ & $+0.88\pm 0.10$ \\
CD$-29^{\circ}$8822   &  5100 & 2.8 & $+0.04\pm 0.15$ & 1.41 &  ... & $1.41\pm 0.30$ & $+0.72\pm 0.35$ & $+0.95\pm 0.10$ \\
CD$-30^{\circ}$8774   &  4900 & 2.3 & $-0.11\pm 0.14$ &  ... &  ... &            ... &             ... & $+0.42\pm 0.11$ \\
CD$-38^{\circ}$585    &  4800 & 2.6 & $-0.52\pm 0.09$ & 1.31 & 1.61 & $1.46\pm 0.25$ & $+1.33\pm 0.30$ & $+1.21\pm 0.11$ \\
CD$-42^{\circ}$2048   &  4400 & 1.6 & $-0.23\pm 0.16$ & 1.41 & 1.31 & $1.36\pm 0.33$ & $+0.94\pm 0.37$ & $+1.04\pm 0.11$ \\
CD$-53^{\circ}$8144   &  4800 & 2.3 & $-0.19\pm 0.15$ & 1.31 & 1.41 & $1.36\pm 0.25$ & $+0.90\pm 0.30$ & $+0.85\pm 0.10$ \\
CD$-61^{\circ}$1941   &  4800 & 2.4 & $-0.20\pm 0.14$ & 1.21 & 1.41 & $1.31\pm 0.25$ & $+0.86\pm 0.30$ & $+0.84\pm 0.11$ \\[0.2 cm]
CPD$-62^{\circ}$1013  &  5100 & 2.6 & $-0.08\pm 0.14$ &  ... &  ... &            ... &             ... & $+0.81\pm 0.11$ \\
CPD$-64^{\circ}$4333  &  4900 & 2.6 & $-0.10\pm 0.18$ & 2.01 & 2.21 & $2.11\pm 0.25$ & $+1.56\pm 0.30$ & $+1.22\pm 0.11$ \\
HD~4084      &  4800 & 2.8 & $-0.42\pm 0.15$ & 1.11 &  ... & $1.11\pm 0.25$ & $+0.88\pm 0.30$ & $+0.81\pm 0.11$ \\
HD~5424      &  4700 & 2.4 & $-0.41\pm 0.18$ & 1.61 & 1.71 & $1.66\pm 0.25$ & $+1.42\pm 0.30$ & $+1.30\pm 0.11$ \\
HD~5825      &  5000 & 2.7 & $-0.48\pm 0.08$ &  ... &  ... &            ... &             ... & $+0.86\pm 0.10$ \\
HD~15589     &  4900 & 3.1 & $-0.27\pm 0.15$ & 1.91 & 1.91 & $1.91\pm 0.25$ & $+1.53\pm 0.30$ & $+1.12\pm 0.11$ \\
HD~20394     &  5100 & 2.9 & $-0.22\pm 0.12$ & 1.61 & 1.81 & $1.71\pm 0.30$ & $+1.28\pm 0.34$ & $+1.15\pm 0.10$ \\
HD~21989     &  4400 & 1.8 & $-0.14\pm 0.17$ & 0.91 & 0.91 & $0.91\pm 0.33$ & $+0.40\pm 0.37$ & $+0.51\pm 0.11$ \\
HD~22285     &  4900 & 2.3 & $-0.60\pm 0.13$ & 1.31 & 1.46 & $1.38\pm 0.25$ & $+1.34\pm 0.30$ & $+1.11\pm 0.10$ \\
HD~22772     &  4800 & 2.4 & $-0.17\pm 0.13$ &  ... &  ... &            ... &             ... & $+0.79\pm 0.11$ \\[0.2 cm]
HD~24035     &  4700 & 2.5 & $-0.23\pm 0.15$ & 1.91 & 1.91 & $1.91\pm 0.25$ & $+1.49\pm 0.30$ & $+1.36\pm 0.11$ \\
HD~29370     &  4800 & 2.1 & $-0.25\pm 0.16$ & 1.61 & 1.71 & $1.66\pm 0.25$ & $+1.26\pm 0.30$ & $+0.87\pm 0.11$ \\
HD~29685     &  4900 & 2.7 & $-0.07\pm 0.14$ &  ... &  ... &            ... &             ... & $+0.55\pm 0.10$ \\
HD~30240     &  5100 & 2.7 & $+0.02\pm 0.15$ &  ... &  ... &            ... &             ... & $+0.66\pm 0.10$ \\
HD~30554     &  4800 & 2.5 & $-0.12\pm 0.14$ & 1.11 &  ... & $1.11\pm 0.25$ & $+0.58\pm 0.30$ & $+0.69\pm 0.10$ \\
HD~32712     &  4600 & 2.1 & $-0.24\pm 0.16$ & 1.31 & 1.41 & $1.36\pm 0.33$ & $+0.95\pm 0.37$ & $+0.82\pm 0.11$ \\
HD~32901     &  4400 & 1.6 & $-0.44\pm 0.14$ & 0.96 & 1.11 & $1.03\pm 0.33$ & $+0.82\pm 0.37$ & $+0.54\pm 0.11$ \\
HD~35993     &  5100 & 2.9 & $-0.05\pm 0.12$ & 1.61 &  ... & $1.61\pm 0.30$ & $+1.01\pm 0.34$ & $+0.99\pm 0.10$ \\
HD~36650     &  4800 & 2.3 & $-0.28\pm 0.13$ &  ... &  ... &            ... &             ... & $+0.57\pm 0.10$ \\
HD~38488     &  4400 & 2.0 & $+0.05\pm 0.10$ & 1.11 & 1.01 & $1.06\pm 0.33$ & $+0.36\pm 0.37$ & $+0.76\pm 0.11$ \\[0.2 cm]
HD~40430     &  4900 & 2.5 & $-0.23\pm 0.13$ &  ... &  ... &            ... &             ... & $+0.71\pm 0.10$ \\
HD~43389     &  4500 & 1.5 & $-0.50\pm 0.17$ & 1.71 & 1.81 & $1.76\pm 0.33$ & $+1.61\pm 0.37$ & $+1.27\pm 0.11$ \\
HD~51959     &  5000 & 3.2 & $-0.10\pm 0.15$ &  ... &  ... &            ... &             ... & $+0.76\pm 0.10$ \\
HD~58368     &  5000 & 2.6 & $+0.04\pm 0.14$ &  ... &  ... &            ... &             ... & $+0.71\pm 0.10$ \\
HD~59852     &  5000 & 2.2 & $-0.22\pm 0.10$ &  ... &  ... &            ... &             ... & $+0.33\pm 0.10$ \\
HD~61332     &  4700 & 2.1 & $+0.07\pm 0.13$ & 1.11 &  ... & $1.11\pm 0.25$ & $+0.39\pm 0.30$ & $+0.40\pm 0.11$ \\
HD~64425     &  4900 & 2.4 & $+0.06\pm 0.16$ & 1.61 & 1.61 & $1.61\pm 0.25$ & $+0.90\pm 0.30$ & $+0.88\pm 0.11$ \\
HD~66291     &  4600 & 1.5 & $-0.31\pm 0.15$ &  ... &  ... &            ... &            ...  & $+0.66\pm 0.11$ \\
HD~67036     &  4300 & 1.5 & $-0.41\pm 0.13$ & 1.06 & 1.11 & $1.09\pm 0.33$ & $+0.85\pm 0.37$ & $+0.90\pm 0.11$ \\
HD~71458     &  4600 & 2.2 & $-0.03\pm 0.10$ & 1.11 & 1.11 & $1.11\pm 0.33$ & $+0.49\pm 0.37$ & $+0.61\pm 0.11$ \\[0.2 cm]
HD~74950     &  4200 & 1.2 & $-0.21\pm 0.13$ & 0.96 & 1.11 & $1.03\pm 0.33$ & $+0.59\pm 0.37$ & $+0.61\pm 0.12$ \\
HD~82221     &  4400 & 1.6 & $-0.21\pm 0.18$ & 1.01 & 1.11 & $1.06\pm 0.33$ & $+0.62\pm 0.37$ & $+0.79\pm 0.11$ \\
HD~83548     &  5000 & 2.4 & $+0.03\pm 0.14$ &  ... &  ... &            ... &            ...  & $+0.62\pm 0.11$ \\
HD~84610     &  4900 & 2.5 & $+0.00\pm 0.14$ &  ... &  ... &            ... &            ...  & $+0.54\pm 0.11$ \\
HD~84678     &  4600 & 1.7 & $-0.13\pm 0.16$ & 1.91 & 1.71 & $1.81\pm 0.33$ & $+1.29\pm 0.37$ & $+1.31\pm 0.11$ \\
HD~88035     &  4900 & 2.4 & $-0.10\pm 0.18$ & 1.46 & 1.71 & $1.59\pm 0.25$ & $+1.04\pm 0.30$ & $+0.95\pm 0.10$ \\
HD~88562     &  4300 & 1.6 & $-0.27\pm 0.15$ &  ... & 1.21 & $1.21\pm 0.33$ & $+0.83\pm 0.37$ & $+0.84\pm 0.11$ \\
HD~89175     &  4900 & 2.1 & $-0.55\pm 0.13$ & 1.61 & 1.91 & $1.76\pm 0.25$ & $+1.66\pm 0.30$ & $+1.36\pm 0.11$ \\
HD~91208     &  5100 & 3.0 & $+0.05\pm 0.14$ &  ... &  ... &            ... &            ...  & $+0.77\pm 0.11$ \\
HD~91979     &  4900 & 2.7 & $-0.11\pm 0.12$ & 1.11 &  ... & $1.11\pm 0.25$ & $+0.57\pm 0.30$ & $+0.80\pm 0.11$ \\[0.2 cm]
HD~92626     &  4800 & 2.3 & $-0.15\pm 0.22$ & 2.21 & 2.36 & $2.28\pm 0.25$ & $+1.78\pm 0.30$ & $+1.38\pm 0.11$ \\
HD~105902    &  4700 & 2.4 & $-0.18\pm 0.17$ & 1.71 & 1.61 & $1.66\pm 0.25$ & $+1.19\pm 0.30$ & $+1.20\pm 0.11$ \\
HD~107264    &  4500 & 1.5 & $-0.19\pm 0.17$ & 1.56 & 1.46 & $1.51\pm 0.33$ & $+1.05\pm 0.37$ & $+0.88\pm 0.11$ \\
HD~107541    &  5000 & 3.2 & $-0.63\pm 0.11$ & 2.01 & 2.15 & $2.08\pm 0.30$ & $+2.06\pm 0.34$ & $+1.59\pm 0.11$ \\
HD~110483    &  4900 & 2.6 & $-0.04\pm 0.14$ & 1.36 &  ... & $1.36\pm 0.25$ & $+0.75\pm 0.30$ & $+0.86\pm 0.11$ \\
HD~110591    &  4700 & 1.8 & $-0.56\pm 0.12$ &  ... &  ... &            ... &             ... & $+0.60\pm 0.10$ \\
HD~111315    &  4900 & 2.0 & $+0.04\pm 0.09$ &  ... &  ... &            ... &             ... & $+0.50\pm 0.11$ \\
HD~113291    &  4700 & 2.6 & $-0.02\pm 0.16$ & 1.46 &  ... & $1.46\pm 0.25$ & $+0.83\pm 0.30$ & $+0.96\pm 0.11$ \\
HD~116869    &  4800 & 2.3 & $-0.36\pm 0.12$ &  ... &  ... &            ... &             ... & $+0.76\pm 0.11$ \\
HD~119185    &  4800 & 2.0 & $-0.43\pm 0.10$ &  ... &  ... &            ... &             ... & $+0.38\pm 0.10$ \\[0.2 cm]
HD~120571    &  4600 & 1.7 & $-0.39\pm 0.09$ & 0.81 & 0.91 & $0.86\pm 0.33$ & $+0.60\pm 0.37$ & $+0.56\pm 0.11$ \\
HD~120620    &  5000 & 3.3 & $-0.14\pm 0.18$ & 2.01 & 2.01 & $2.01\pm 0.30$ & $+1.50\pm 0.35$ & $+1.37\pm 0.10$ \\
HD~122687    &  5000 & 2.6 & $-0.07\pm 0.13$ & 1.31 &  ... & $1.31\pm 0.30$ & $+0.73\pm 0.35$ & $+0.93\pm 0.10$ \\
HD~123396    &  4600 & 1.9 & $-1.04\pm 0.13$ & 0.81 & 1.21 & $1.01\pm 0.33$ & $+1.40\pm 0.37$ & $+0.98\pm 0.10$ \\
HD~123701    &  5000 & 2.5 & $-0.44\pm 0.09$ & 1.31 & 1.61 & $1.46\pm 0.30$ & $+1.25\pm 0.34$ & $+1.04\pm 0.10$ \\
HD~123949    &  4600 & 2.2 & $-0.09\pm 0.18$ & 1.86 & 1.81 & $1.84\pm 0.33$ & $+1.28\pm 0.37$ & $+1.21\pm 0.11$ \\
HD~126313    &  4900 & 2.2 & $-0.10\pm 0.16$ & 1.36 & 1.71 & $1.54\pm 0.25$ & $+0.99\pm 0.30$ & $+0.89\pm 0.11$ \\
HD~130255    &  4400 & 1.5 & $-1.11\pm 0.11$ &  ... &  ... &            ... &             ... & $+0.34\pm 0.11$ \\
HD~131670    &  4700 & 2.2 & $-0.04\pm 0.15$ & 1.06 &  ... & $1.06\pm 0.25$ & $+0.45\pm 0.30$ & $+0.58\pm 0.11$ \\
HD~136636    &  4900 & 2.5 & $-0.04\pm 0.18$ & 1.31 & 1.11 & $1.21\pm 0.25$ & $+0.60\pm 0.30$ & $+0.94\pm 0.10$ \\[0.2 cm]
HD~142751    &  4600 & 2.0 & $-0.10\pm 0.13$ & 1.06 & 1.11 & $1.09\pm 0.33$ & $+0.54\pm 0.37$ & $+0.69\pm 0.11$ \\
HD~143899    &  5000 & 2.5 & $-0.27\pm 0.12$ &  ... &  ... &            ... &             ... & $+0.72\pm 0.10$ \\
HD~147884    &  5100 & 3.0 & $-0.09\pm 0.15$ &  ... &  ... &            ... &             ... & $+0.86\pm 0.10$ \\
HD~148177    &  4400 & 1.6 & $-0.15\pm 0.15$ & 0.61 & 0.81 & $0.71\pm 0.33$ & $+0.21\pm 0.37$ & $+0.58\pm 0.11$ \\
HD~154430    &  4200 & 1.2 & $-0.36\pm 0.19$ & 1.51 & 1.21 & $1.36\pm 0.33$ & $+1.07\pm 0.37$ & $+1.07\pm 0.11$ \\
HD~162806    &  4500 & 1.7 & $-0.26\pm 0.17$ & 0.96 & 1.11 & $1.03\pm 0.33$ & $+0.64\pm 0.37$ & $+0.78\pm 0.11$ \\
HD~168214    &  5300 & 3.2 & $-0.08\pm 0.10$ &  ... &  ... &            ... &             ... & $+0.92\pm 0.10$ \\
HD~168560    &  4400 & 1.6 & $-0.13\pm 0.13$ & 0.86 & 1.01 & $0.94\pm 0.33$ & $+0.42\pm 0.37$ & $+0.45\pm 0.11$ \\
HD~168791    &  4400 & 1.7 & $-0.23\pm 0.17$ & 1.31 & 1.41 & $1.36\pm 0.33$ & $+0.94\pm 0.37$ & $+0.81\pm 0.11$ \\
HD~176105    &  4500 & 1.6 & $-0.14\pm 0.12$ & 0.76 & 1.01 & $0.88\pm 0.33$ & $+0.38\pm 0.37$ & $+0.48\pm 0.11$ \\[0.2 cm]
HD~177192    &  4700 & 1.7 & $-0.17\pm 0.20$ &  ... &  ... &            ... &             ... & $+0.53\pm 0.10$ \\
HD~180996    &  4900 & 2.6 & $+0.06\pm 0.15$ &  ... &  ... &            ... &             ... & $+0.52\pm 0.11$ \\
HD~182300    &  5000 & 2.7 & $+0.06\pm 0.16$ & 1.31 &  ... & $1.31\pm 0.30$ & $+0.60\pm 0.35$ & $+0.87\pm 0.10$ \\
HD~183915    &  4500 & 1.6 & $-0.39\pm 0.14$ & 1.31 & 1.36 & $1.34\pm 0.33$ & $+1.08\pm 0.37$ & $+1.00\pm 0.11$ \\
HD~187308    &  4900 & 2.5 & $-0.08\pm 0.11$ &  ... &  ... &            ... &             ... & $+0.63\pm 0.11$ \\
HD~193530    &  4400 & 1.6 & $-0.17\pm 0.14$ &  ... &  ... &            ... &             ... & $+0.57\pm 0.11$ \\
HD~196445    &  4400 & 1.4 & $-0.19\pm 0.17$ & 1.51 & 1.41 & $1.46\pm 0.33$ & $+1.00\pm 0.37$ & $+1.03\pm 0.11$ \\
HD~199435    &  5000 & 2.6 & $-0.39\pm 0.12$ & 1.41 & 1.71 & $1.56\pm 0.30$ & $+1.30\pm 0.34$ & $+1.07\pm 0.10$ \\
HD~200995    &  4600 & 2.1 & $-0.03\pm 0.17$ & 1.06 & 1.11 & $1.09\pm 0.33$ & $+0.47\pm 0.37$ & $+0.62\pm 0.11$ \\
HD~201657    &  4700 & 2.2 & $-0.34\pm 0.17$ & 1.56 & 1.61 & $1.59\pm 0.25$ & $+1.28\pm 0.30$ & $+1.12\pm 0.11$ \\[0.2 cm]
HD~201824    &  4900 & 2.3 & $-0.33\pm 0.17$ & 1.41 & 1.76 & $1.59\pm 0.25$ & $+1.27\pm 0.30$ & $+1.14\pm 0.11$ \\
HD~204075    &  5300 & 1.5 & $+0.06\pm 0.17$ &  ... &  ... &            ... &             ... & $+0.89\pm 0.10$ \\
HD~207277    &  4600 & 2.0 & $-0.13\pm 0.14$ & 1.11 & 1.11 & $1.11\pm 0.33$ & $+0.59\pm 0.37$ & $+0.72\pm 0.11$ \\
HD~210709    &  4700 & 2.6 & $-0.10\pm 0.14$ & 0.91 &  ... & $0.91\pm 0.25$ & $+0.36\pm 0.30$ & $+0.60\pm 0.11$ \\
HD~210946    &  4800 & 2.1 & $-0.12\pm 0.13$ &  ... &  ... &            ... &             ... & $+0.63\pm 0.11$ \\
HD~211173    &  4700 & 2.3 & $-0.39\pm 0.09$ &  ... &  ... &            ... &             ... & $+0.45\pm 0.10$ \\
HD~211594    &  4900 & 2.4 & $-0.43\pm 0.14$ & 1.61 & 1.76 & $1.68\pm 0.25$ & $+1.46\pm 0.30$ & $+1.35\pm 0.10$ \\
HD~211954    &  4400 & 1.7 & $-0.51\pm 0.19$ & 1.91 & 1.71 & $1.81\pm 0.33$ & $+1.67\pm 0.37$ & $+1.28\pm 0.11$ \\
HD~214579    &  4400 & 1.6 & $-0.26\pm 0.14$ & 1.11 & 1.11 & $1.11\pm 0.33$ & $+0.72\pm 0.37$ & $+0.70\pm 0.11$ \\
HD~217143    &  4600 & 2.1 & $-0.35\pm 0.17$ & 1.11 & 1.11 & $1.11\pm 0.33$ & $+0.81\pm 0.37$ & $+0.88\pm 0.11$ \\[0.2 cm]
HD~217447    &  5000 & 2.5 & $-0.17\pm 0.11$ &  ... &  ... &            ... &             ... & $+0.85\pm 0.10$ \\
HD~219116    &  4900 & 2.3 & $-0.61\pm 0.09$ &  ... &  ... &            ... &             ... & $+0.85\pm 0.11$ \\
HD~223586    &  4700 & 2.9 & $-0.08\pm 0.11$ & 1.21 &  ... & $1.21\pm 0.25$ & $+0.64\pm 0.30$ & $+0.83\pm 0.11$ \\
HD~223617    &  4700 & 2.3 & $-0.18\pm 0.13$ & 1.01 &  ... & $1.01\pm 0.25$ & $+0.54\pm 0.30$ & $+0.67\pm 0.11$ \\
HD~252117    &  4600 & 1.8 & $-0.14\pm 0.19$ & 1.41 & 1.41 & $1.41\pm 0.33$ & $+0.90\pm 0.37$ & $+0.95\pm 0.11$ \\
HD~273845    &  4800 & 2.5 & $-0.15\pm 0.16$ & 1.61 & 1.61 & $1.61\pm 0.25$ & $+1.11\pm 0.30$ & $+1.03\pm 0.10$ \\
HD~288174    &  4800 & 2.3 & $-0.05\pm 0.15$ & 1.21 &  ... & $1.21\pm 0.25$ & $+0.61\pm 0.30$ & $+0.68\pm 0.10$ \\
MFU~112      &  4900 & 2.4 & $-0.43\pm 0.15$ & 1.86 & 2.06 & $1.96\pm 0.25$ & $+1.74\pm 0.30$ & $+1.33\pm 0.11$ \\
BD$-18^{\circ}$821     &  5000 & 2.3 & $-0.27\pm 0.15$ & 1.46 & 1.61 & $1.54\pm 0.30$ & $+1.16\pm 0.35$ & $+0.87\pm 0.11$ \\
CD$-26^{\circ}$7844    &  5100 & 2.8 & $+0.02\pm 0.11$ &  ... &  ... &            ... &             ... & $+0.45\pm 0.10$ \\[0.2 cm]
CD$-30^{\circ}$9005    &  4700 & 2.3 & $+0.05\pm 0.12$ & 1.21 & 1.11 & $1.16\pm 0.25$ & $+0.46\pm 0.30$ & $+0.62\pm 0.10$ \\
CD$-34^{\circ}$6139    &  4900 & 2.2 & $-0.07\pm 0.13$ &  ... &  ... &            ... &             ... & $+0.58\pm 0.11$ \\
CD$-34^{\circ}$7430    &  4900 & 2.6 & $+0.01\pm 0.14$ &  ... &  ... &            ... &             ... & $+0.55\pm 0.11$ \\
CD$-46^{\circ}$3977    &  4900 & 2.6 & $-0.10\pm 0.15$ &  ... &  ... &            ... &             ... & $+0.65\pm 0.10$ \\
HD~18182     &  4900 & 2.4 & $-0.17\pm 0.10$ &  ... &  ... &            ... &             ... & $+0.41\pm 0.11$ \\
HD~18361     &  4900 & 2.6 & $+0.01\pm 0.15$ &  ... &  ... &            ... &             ... & $+0.40\pm 0.10$ \\
HD~21682     &  5200 & 2.8 & $-0.48\pm 0.12$ &  ... &  ... &            ... &             ... & $+0.66\pm 0.10$ \\
HD~26886     &  5000 & 2.5 & $-0.30\pm 0.10$ &  ... &  ... &            ... &             ... & $+0.56\pm 0.10$ \\
HD~31812     &  5100 & 2.6 & $-0.07\pm 0.11$ &  ... &  ... &            ... &             ... & $+0.58\pm 0.10$ \\
HD~33709     &  5000 & 2.1 & $-0.20\pm 0.14$ &  ... &  ... &            ... &             ... & $+0.39\pm 0.10$ \\[0.2 cm]
HD~39778     &  5000 & 2.5 & $-0.12\pm 0.12$ &  ... &  ... &            ... &             ... & $+0.84\pm 0.10$ \\
HD~41701     &  5000 & 2.6 & $+0.02\pm 0.13$ &  ... &  ... &            ... &             ... & $+0.34\pm 0.10$ \\
HD~45483     &  4800 & 2.2 & $-0.14\pm 0.12$ &  ... &  ... &            ... &             ... & $+0.57\pm 0.10$ \\
HD~48814     &  4800 & 2.3 & $-0.07\pm 0.11$ &  ... &  ... &            ... &             ... & $+0.32\pm 0.10$ \\
HD~49017     &  5100 & 2.8 & $+0.02\pm 0.11$ &  ... &  ... &            ... &             ... & $+0.24\pm 0.10$ \\
HD~49661     &  5000 & 2.4 & $-0.13\pm 0.10$ &  ... &  ... &            ... &             ... & $+0.24\pm 0.10$ \\
HD~49778     &  5000 & 2.3 & $-0.22\pm 0.12$ &  ... &  ... &            ... &             ... & $+0.23\pm 0.10$ \\
HD~50075     &  4900 & 2.5 & $-0.16\pm 0.11$ & 1.16 &  ... & $1.16\pm 0.25$ & $+0.67\pm 0.30$ & $+0.72\pm 0.10$ \\
HD~50843     &  4700 & 2.3 & $-0.31\pm 0.13$ & 0.61 &  ... & $0.61\pm 0.25$ & $+0.27\pm 0.30$ & $+0.44\pm 0.11$ \\
HD~53199     &  5000 & 2.3 & $-0.23\pm 0.13$ &  ... &  ... &            ... &             ... & $+0.77\pm 0.10$ \\[0.2 cm]
HD~58121     &  4600 & 1.8 & $-0.01\pm 0.13$ &  ... &  ... &            ... &             ... & $+0.30\pm 0.11$ \\
HD~88495     &  4900 & 3.0 & $-0.11\pm 0.10$ &  ... &  ... &            ... &             ... & $+0.67\pm 0.11$ \\
HD~90167     &  5000 & 2.6 & $-0.04\pm 0.11$ &  ... &  ... &            ... &             ... & $+0.40\pm 0.11$ \\
HD~95193     &  5000 & 2.7 & $+0.04\pm 0.12$ &  ... &  ... &            ... &             ... & $+0.52\pm 0.10$ \\
HD~107270    &  5400 & 2.7 & $+0.05\pm 0.17$ &  ... &  ... &            ... &             ... & $+0.41\pm 0.10$ \\
HD~109061    &  4700 & 2.0 & $-0.56\pm 0.09$ &  ... &  ... &            ... &             ... & $+0.62\pm 0.10$ \\
HD~113195    &  4700 & 2.1 & $-0.15\pm 0.12$ &  ... &  ... &            ... &             ... & $+0.50\pm 0.11$ \\
HD~115277    &  4800 & 2.4 & $-0.03\pm 0.15$ &  ... &  ... &            ... &             ... & $+0.41\pm 0.11$ \\
HD~119650    &  4500 & 1.6 & $-0.10\pm 0.13$ &  ... &  ... &            ... &             ... & $+0.27\pm 0.11$ \\
HD~134698    &  4500 & 1.7 & $-0.52\pm 0.12$ &  ... &  ... &            ... &             ... & $+0.49\pm 0.11$ \\[0.2 cm]
HD~139266    &  4300 & 1.5 & $-0.27\pm 0.18$ & 1.11 & 1.11 & $1.11\pm 0.33$ & $+0.73\pm 0.37$ & $+0.81\pm 0.11$ \\
HD~139409    &  4700 & 2.1 & $-0.51\pm 0.13$ &  ... &  ... &            ... &             ... & $+0.47\pm 0.10$ \\
HD~169106    &  4900 & 2.2 & $+0.01\pm 0.12$ &  ... &  ... &            ... &             ... & $+0.36\pm 0.10$ \\
HD~184001    &  5000 & 2.5 & $-0.21\pm 0.14$ &  ... &  ... &            ... &             ... & $+0.59\pm 0.11$ \\
HD~204886    &  4600 & 2.1 & $+0.04\pm 0.15$ & 1.31 & 1.26 & $1.28\pm 0.33$ & $+0.59\pm 0.37$ & $+0.66\pm 0.16$ \\
HD~213084    &  5000 & 2.8 & $-0.09\pm 0.15$ & 1.21 &  ... & $1.21\pm 0.30$ & $+0.65\pm 0.35$ & $+0.81\pm 0.10$ \\
HD~223938    &  5000 & 2.9 & $-0.42\pm 0.11$ &  ... &  ... &            ... &             ... & $+0.74\pm 0.10$ \\
MFU~214      &  4800 & 2.4 & $+0.00\pm 0.12$ &  ... &  ... &            ... &             ... & $+0.30\pm 0.11$ \\
MFU~229      &  4900 & 2.6 & $-0.01\pm 0.11$ &  ... &  ... &            ... &             ... & $+0.54\pm 0.10$ \\
HD~12392     &  4900 & 3.0 & $-0.08\pm 0.18$ & 1.71 & 1.81 & $1.76\pm 0.25$ & $+1.19\pm 0.30$ & $+1.14\pm 0.11$ \\[0.2 cm]
HD~17067     &  4200 & 1.1 & $-0.61\pm 0.21$ & 1.26 & 1.31 & $1.28\pm 0.33$ & $+1.25\pm 0.37$ & $+0.94\pm 0.11$ \\
HD~90127     &  4600 & 2.2 & $-0.40\pm 0.10$ & 1.06 &  ... & $1.06\pm 0.33$ & $+0.81\pm 0.37$ & $+0.79\pm 0.11$ \\
HD~102762    &  4400 & 1.7 & $-0.17\pm 0.20$ & 1.66 & 1.61 & $1.63\pm 0.33$ & $+1.15\pm 0.37$ & $+1.01\pm 0.11$ \\
HD~114678    &  5200 & 2.8 & $-0.50\pm 0.13$ & 1.91 & 1.71 & $1.81\pm 0.30$ & $+1.66\pm 0.34$ & $+1.17\pm 0.10$ \\
HD~180622    &  4600 & 2.2 & $+0.03\pm 0.12$ & 0.81 &  ... & $0.81\pm 0.33$ & $+0.13\pm 0.37$ & $+0.43\pm 0.11$ \\
HD~200063    &  4100 & 1.1 & $-0.34\pm 0.20$ & 1.16 & 1.16 & $1.16\pm 0.33$ & $+0.85\pm 0.37$ & $+0.83\pm 0.11$ \\
HD~210030    &  4700 & 1.9 & $-0.03\pm 0.11$ &  ... &  ... &            ... &             ... & $+0.31\pm 0.11$ \\
HD~214889    &  4900 & 2.6 & $-0.17\pm 0.12$ &  ... &  ... &            ... &             ... & $+0.58\pm 0.10$ \\
HD~215555    &  5200 & 3.2 & $-0.08\pm 0.12$ &  ... &  ... &            ... &             ... & $+0.84\pm 0.10$ \\
HD~216809    &  4400 & 1.2 & $-0.04\pm 0.14$ & 0.91 & 1.11 & $1.01\pm 0.33$ & $+0.40\pm 0.37$ & $+0.49\pm 0.11$ \\[0.2 cm]
HD~221879    &  4300 & 1.3 & $-0.10\pm 0.19$ & 1.01 & 1.11 & $1.06\pm 0.33$ & $+0.51\pm 0.37$ & $+0.66\pm 0.11$ \\
HD~749       &  4700 & 2.6 & $-0.29\pm 0.15$ & 1.16 & 1.26 & $1.21\pm 0.25$ & $+0.85\pm 0.30$ & $+0.89\pm 0.11$ \\
HD~88927     &  4600 & 2.3 & $+0.02\pm 0.13$ &  ... &  ... &            ... &             ... & $+0.33\pm 0.11$ \\
BD$+09^{\circ}$2384   &  4900 & 2.5 & $-0.98\pm 0.10$ &  ... &  ... &            ... &            ... & $+0.72\pm 0.11$ \\
HD~89638     &  4900 & 2.4 & $-0.19\pm 0.11$ &  ... &  ... &            ... &             ... & $+0.57\pm 0.10$ \\
HD~187762    &  4800 & 2.4 & $-0.30\pm 0.11$ &  ... &  ... &            ... &             ... & $+0.40\pm 0.11$ \\
NGC~5822-201 &  5200 & 2.7 & $-0.11\pm 0.10$ &  ... &  ... &            ... &             ... & $+0.80\pm 0.10$ \\
NGC~5822-2   &  5100 & 2.4 & $-0.15\pm 0.09$ &  ... &  ... &            ... &             ... & $+0.75\pm 0.10$ \\
HD~10613     &  4950 & 2.7 & $-0.92\pm 0.12$ & 1.11 & 1.11 & $1.11\pm 0.25$ & $+1.38\pm 0.30$ & $+1.25\pm 0.11$ \\
CD$-25^{\circ}$6606   &  5300 & 2.7 & $+0.12\pm 0.14$ &  ... &  ... &            ... &            ... & $+0.58\pm 0.11$ \\[0.2 cm]
HD~46040     &  4800 & 2.4 &  $+0.11\pm 0.13$ & 1.61 & 1.61 & $1.61\pm 0.25$ & $+0.83\pm 0.30$ & $+0.97\pm 0.11$ \\
HD~49841     &  5200 & 3.2 &  $+0.21\pm 0.13$ &  ... &  ... &            ... &             ... & $+0.68\pm 0.10$ \\
HD~82765     &  5100 & 2.6 &  $+0.19\pm 0.10$ &  ... &  ... &            ... &             ... & $+0.30\pm 0.11$ \\
HD~84734     &  5200 & 2.9 &  $+0.20\pm 0.12$ &  ... &  ... &            ... &             ... & $+0.60\pm 0.10$ \\
HD~85205     &  5300 & 2.8 &  $+0.23\pm 0.16$ &  ... &  ... &            ... &             ... & $+0.60\pm 0.11$ \\
HD~101079    &  5000 & 2.7 &  $+0.10\pm 0.12$ &  ... &  ... &            ... &             ... & $+0.46\pm 0.10$ \\
HD~130386    &  4900 & 2.7 &  $+0.16\pm 0.13$ &  ... &  ... &            ... &             ... & $+0.48\pm 0.11$ \\
HD~139660    &  5000 & 2.8 &  $+0.26\pm 0.14$ &  ... &  ... &            ... &             ... & $+0.47\pm 0.11$ \\
HD~198590    &  5100 & 2.6 &  $+0.18\pm 0.14$ &  ... &  ... &            ... &             ... & $+0.42\pm 0.10$ \\
HD~212209    &  4700 & 2.4 &  $+0.30\pm 0.13$ &  ... &  ... &            ... &             ... & $+0.25\pm 0.11$ \\
\end{longtable}

%%%%%%%%%%%%%%%%%%%%%%%%%%%%%%%%%%%%%%%%%%%%%%%%%%

% Don't change these lines
\bsp	% typesetting comment
\label{lastpage}
\end{document}